# Chirality as an Instrument of Stratification of Hierarchical Systems in Animate and Inanimate Nature

## Vsevolod A. Tverdislov

Faculty of Physics, Lomonosov Moscow State University
*Russia, 119991, Moscow, GSP-1, 1-2 Leninskiye Gory*
*Email: tverdislov@mail.ru*

The article seeks to formulate a synergetic law that is posited to be of common physicochemical and biological nature: an evolving system, possessing free energy and elements with chiral asymmetry may change the type of symmetry inside one hierarchical level, thereby increasing its «complexity», but preserving the sign of predominant chirality («right»-D or «left»-L twist). The same system has a tendency to spontaneous formation of a succession of hierarchical levels with alternating chirality sign of de-novo formed structures and with an increase of the structures' relative scale.

In the living systems the hierarchy principle of conjugated levels of macromolecular structures, starting with the «lower» level of asymmetrical carbon, serves as an anti-entropic factor and also as the structural basis of the «selected mechanical degrees of freedom» in the molecular machines. Observations present evidence of regular alternations of the chirality sign D-L-D-L and L-D-L-D for DNA and protein structures, respectively, during the transition of DNA and proteins to a higher level of structural and functional organization.

Sign-alternating chiral hierarchies of DNA and proteins, in turn, form a complimentary conjugated pair that is an achiral invariant, which closes a molecular biological module of living systems.

The ability of a carbon atom to form chiral compounds is a significant factor that defined carbon basis of living systems on Earth, and also their development via succession of chiral bifurcations. Restricted by the chirality sign, the hierarchy principle of macromolecular structures determined the «modular» character of biological evolution.

*Key words: symmetry, chirality, hierarchy principles, DNA, proteins, fractals*

## Introduction

It is hardly to be expected that the «road map» and driving forces of biological evolution could be understood exclusively in the frame of biology, without the recourse to the general concepts of physics and chemistry. A paradigm of «end-to-end evolution» implies the presence of universal laws and general mechanisms, which permeate the non-living and living nature in the process of its development. Following theoretical physics with its cardinal geometrization of the models of matter, space and time, which use representations of superstrings, joined circular formations, and spirals (Penrose 2005), there will necessarily come a stage of geometrization in theoretical biology. Up till now biology has known analysis of two types of patterns: chemical compounds and morphological structures of cells, organs, tissues and also those of ecosystems. This is a level of descriptive classification. It is of high necessity to identify those basic geometric representations, which would comprise more general spatio-temporal ideas about the structure and evolution of living systems.

Ideas, based on the evolution of the symmetrical states of a system, comprise one of the most adequate approaches. The principle of a modular, hierarchically structured biological evolution naturally follows from similar theoretical considerations.

Unity and struggle of the opposites as a source of the generalized driving force behind the development of systems presuppose the presence of interacting antipodes. Beyond all doubt, the most perfect dialectic pair is the chiral pair of objects or processes, which are of the same nature and dimension, but not principally reducible to each other. This might be the reason for the Nature to have employed the chirality principle as one of the basic principles underlying the transition from non-living to living systems.

The present manuscript covers the development of ideas about chirality as the primary switching instrument in the chain of symmetrical states of evolving systems, initially abiogenic, and later biological. We use the classical definition of chirality given by Lord Kelvin in 1904: «I call any geometrical figure, or group of points, chiral, and say it has chirality, if its image in a plane mirror, ideally realized, cannot be brought to coincide with itself» (Kelvin 1904). More specifically, this article deals with the changes of scale and, more importantly, changes of chirality sign («right» or «left» twist) during a transition between levels in hierarchical systems.

We propose to formulate a new synergetic law that, in our view, is of common physicochemical and biological nature: **in the process of self-organization an evolving system, possessing free energy and elements with chiral asymmetry, may change the type of symmetry inside one hierarchical level, thereby increasing its «complexity», but with transition to a higher level the system may also change its scale and chirality sign together with a change in the functionality of an enantiomorph (Tverdislov et al. 2012)**.

The change of chirality sign provides for the evolutionary irreversibility of stratification (stratification is understood as the splitting of a hierarchical level of a system into two levels with different chirality signs, i.e. formation of chiral layers in a system). Indeed, to switch its chirality sign a chiral object, which does not possess a symmetry axis, must be disassembled and then reassembled from the same elements. The chirality of biological structures of different levels ensures that the process of L/D stratification is universal (for the chirality sign notation we use a generalized Fisher's classification (Buxton and Roberts 1997)), and hierarchical levels are deterministic and stable. An enantiomorph of a higher level, while preserving its mirror equivalence, acquires a broader spectrum of functionality. The hierarchical principle, in turn, uniquely determines the expansion vector of evolving biological systems towards open boundaries – strata.

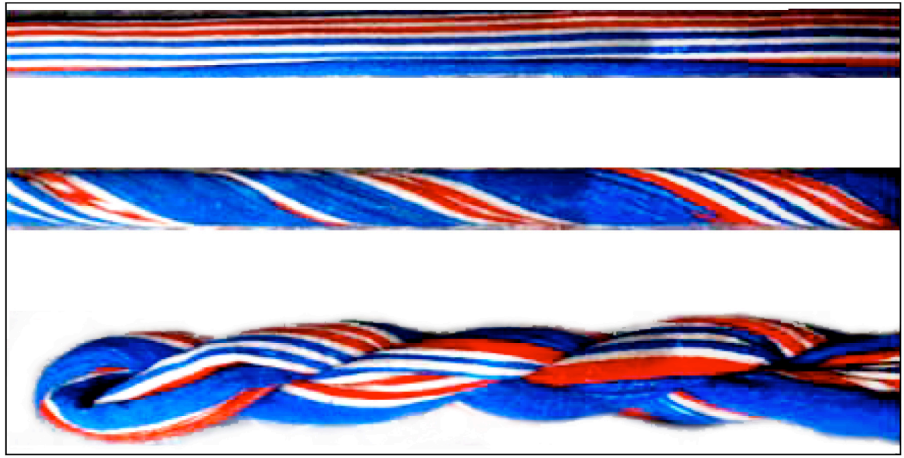

**Fig. 1.** Elastic cord, twisted into a «right» spiral, when folded, forms a «left» superhelix.

## 1. Stratification in chiral abiogenic systems

Consider that various pre-biological systems, physical and physicochemical, can spontaneously change their chirality sign during transition between separate thermodynamic levels (also, during phase transitions) in equilibrium systems or when crossing a bifurcation point in self-organization processes in dissipative systems (in distributed active media, in particular). Here are some examples:

**Twisting of elastic threads.** An elastic cord, twisted with excessive tension into a single spiral (L or D), if folded, spontaneously forms a secondary superhelix of opposite sign (D or L, correspondingly), where the pitch and the radius of a new spiral are increased (Fig. 1). In this case the system lowers its free energy due to a decrease of elastic tension and redistribution of the released energy to new degrees of freedom. The same effect is observed when two parallel cords are twisted around each other. Similar processes happen during supercoiling of circular DNA (Waigh 2007; Nelson 2012).

**Convective structures on the surface of a liquid.** Hizhnyak et al. (Ivanitskii et al. 2005) described emergence of spiral convective microstructures on the surface of a cooling or evaporating liquid. Thermograms record their origination as a result of superposition of the thermo-gravitational effect of Rayleigh–Bénard convection and the capillar effect of Marangoni (Fig. 2). It is essential to note the cooperative surface effects in subsurface temperature gradient: when microflows are curling clockwise, the whole surface layer is turning into the opposite direction. Here the law of conservation of angular momentum is observed.

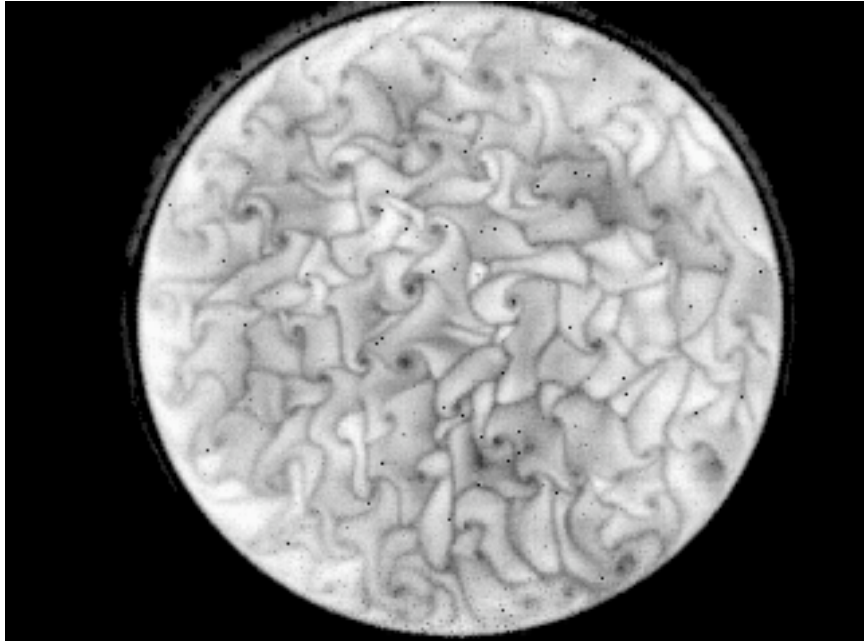

Fig. 2. Emergence of convective structures on the surface of a cooling liquid (40-60C). Structures are visualized in the infrared light. The structures have «right» direction of vortices because the liquid, cooled as a result of evaporation and thermoexchange, descends, curling to the «right». The whole surface layer is turning to the "left" counterclockwise (Ivanitskii et al. 2005).

**Langmuir textures on the surface of chiral solutions.** An illustrative example of changing a chirality sign together with the scale of a structure is the formation of quasi-stationary helical structures on the surface of water solutions of synthetic L- and D-phospholipids (Nandi and Vollhardt 2003). Using Brewster angle microscopy it has been shown that «left» isomers form «right» helices of several millimeters in size, and «right» isomers form «left» «helices». (Fig. 3). Thus, of molecular asymmetry changes to asymmetry of macroscopic scale (that of two-dimensional crystals), which is, essentially, of a different sign of chirality. Note that the surface structure (texture), formed from a racemic solution, includes both types of spirals.

**Chiral strings and superspirals.** Another example is drawn from the works of Stovbun and colleagues (Stovbun and Skoblin 2012). It has been shown experimentally that in homochiral solutions of trifluoroacetylated aminoalcohols in cyclohexane, benzol and other solvents, and also in water solution of phenylalanine, one can observe formation of strings: anisometric (ratio of length to diameter is about $10^2 - 10^5$) helical structures with a characteristic stiffness (Fig. 4a). Due to dipole-dipole interactions, molecules form supramolecular chiral strings, twisted in superspirals that successively change the direction of spirallization. A semi-empirical rule, formulated earlier by these authors, of changing a chirality sign during transition to structures of higher level has been confirmed. A remarkable peculiarity of this bio-mimetic system is the ability of the strings to form twisted double-helices and the ability of the folded strings to form superhelices of the opposite sign (Fig. 4b).

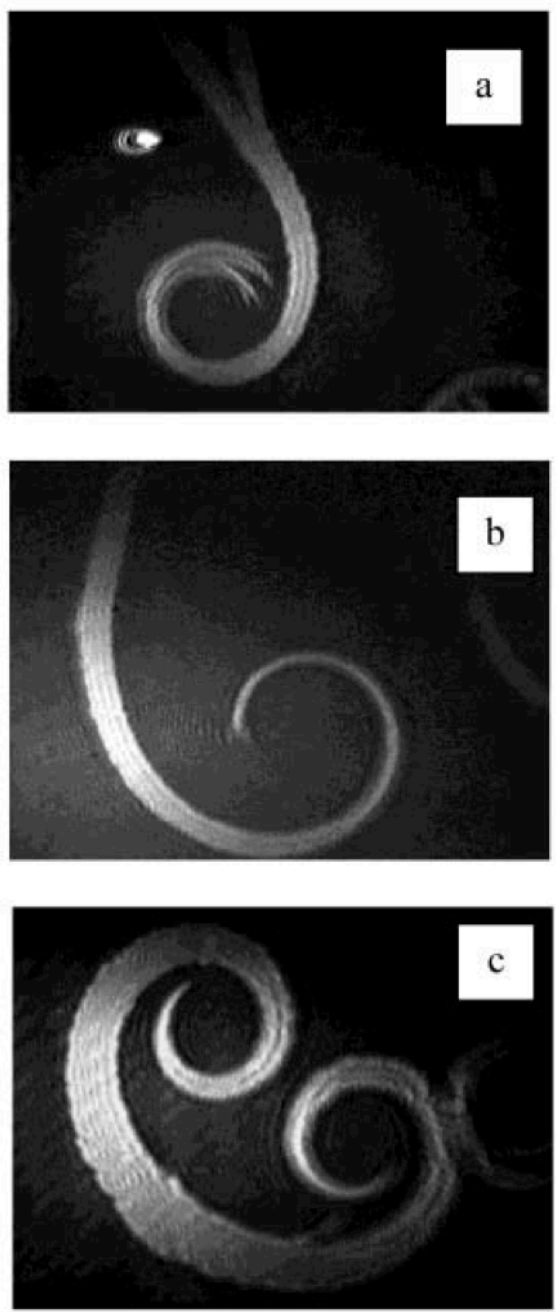

Fig. 3. Domains of condensed phase, which are formed in Langmuir monolayers of enantiomers and racemate of N-R-palmitoyl-threonine, have a spiral form:
(a) – D-enantiomers form «left» spiral;
(b) – L-enantiomers form «right» spiral;
(c) – racemate causes superposition of spirals.
Dimension of images obtained using Brewster angle microscopy (He-Ne laser) - 350 × 350 mkm (Nandi and Vollhardt 2003).

## 2. Carbon as a key element of stratification in living systems

A common view of carbon as «the basis of life» has a foundation in its properties, «suitable» to form robust but labile chemical bonds, which facilitate formation of a large number of linear, branched, cyclic carbon skeletons, providing for a great variety of biological molecules (Waigh 2007). Another property of carbon, crucial for the biosphere, is its ability to form water-soluble and volatile compounds. However, in our view, an even more important biogenic property of carbon is its ability to form chiral compounds that allows for dualism – the primary basis of evolutionary formation of complex hierarchical systems.

As it has been known from numerous publications following the pioneer work of Stanley Miller, all simple biomacromolecule-constituting carbon compounds could spontaneously originate from gases in the Earth's earlier atmosphere under the action of ultraviolet light, electrical discharge, radiation or under other conditions (Lazcano and Bada 2003). For thermodynamic reasons the chiral compounds originated in the achiral non-enantiospecific media in the form of racemates.

For creation of informationally determined macromolecular cellular structures (DNA, RNA, proteins) the Nature used one-dimensional non-branched polymers, maximally «adapted» to recording and reading of information as well as to the ergonomic synthesis, to the «expedient» steric packaging, and also to the conformational functional reconstructions. Critically important non-ambiguity of information recording, storage and reading, disrupted by potential ambiguity of chiral monomeres, is provided for in the cell by the stereospecific enzymatic and ribosomal synthesis (Tverdislov and Yakovenko 2008). The process of primary separation of abiogenically emerging enantiomers, as well as that of establishing homochirality of the proteins and nuclear acids should be attributed, as it seems, to the origins of biopoesis, i.e. the pre-biological or early biological stages, when proto-cells were in the stage of formation. During the pre-biological stage this process could have been caused with the homochiral linear polimerization (during two-dimensional cristallization at the boondary of phase separation) in pre-biotic physicochemical systems (Gol'danskii and Kuz'min 1989); during the early biological stage this process might have been connected with the discard of enantiomers of one sign in the course of biological selection (Barron 2009). In principle, the initial selection of an enantiomer's sign could be random.

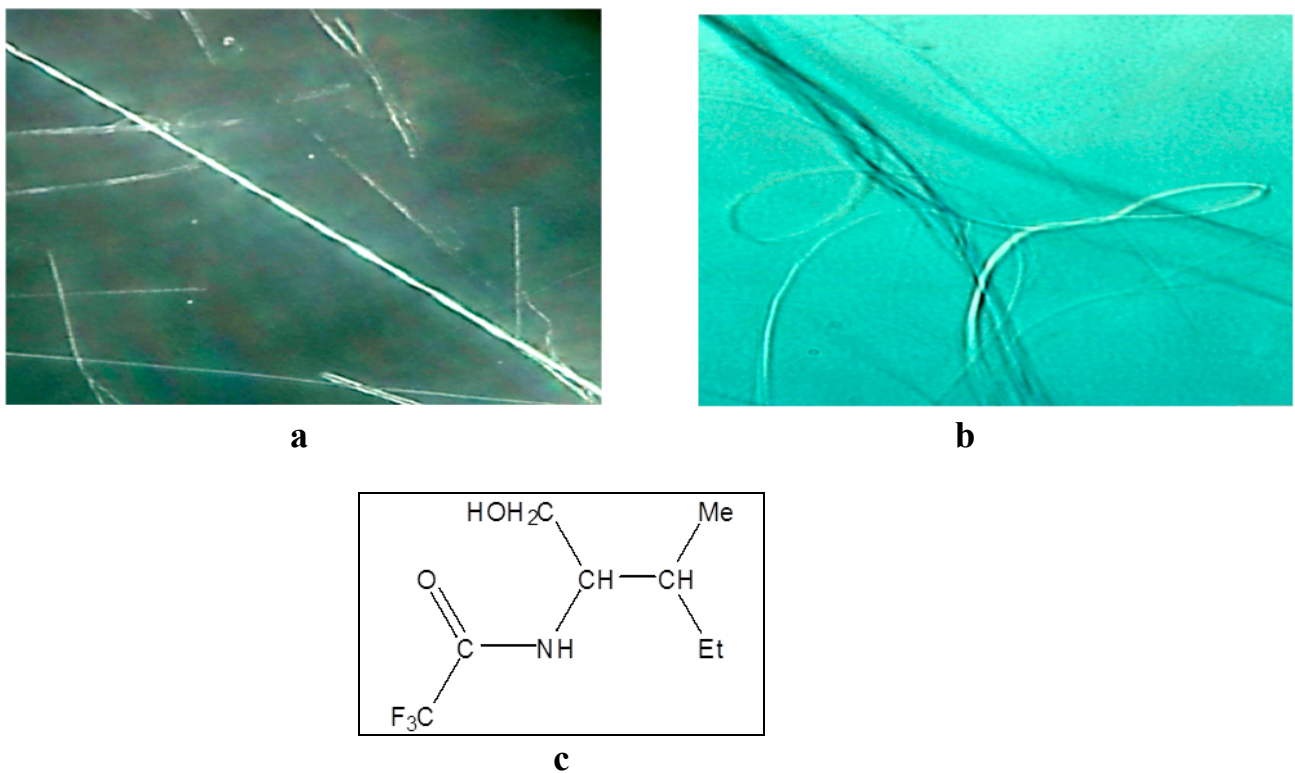

Fig. 4. Double «left» spiral that emerged from the «right-hand twisted» strings (a) and «right» superspiral that emerged from the «left» spirals (b). Spirals are formed from L-homochiral solution of trifluoroacetylated aminoalcohol (c).

To believe that discrimination of primary chiral bioorganic substances was an autonomous process is at least undeliberate. From the biological point of view, there exists a certain minimal set of physicochemical characteristics, which probionts, the precursors of living cells, should have possessed. These are discreteness, thermodynamical non-equilibrium, a certain ionic composition, which mirrors that of seawater, chiral purity of proteins, sugars and lipids. It seems that following the origination of primary vesicular physicochemical reactors, an apparatus of covariant matrix synthesis was formed, and abilities for self-reproduction, excitability and motility emerged. The mechanism of origination of all these properties as one complex remains a mystery. In principle, all of these properties except for the chiral purity could have been realized **separately** in different parts of the ancient geosphere: in the estuaries of world oceans, in the «Darwinian puddles», in geothermal sources, on the surface of optically active crystals, in polarized electromagnetic fields (Tverdislov and Yakovenko 2008). This subject has been amply commented on in the literature. Homochirality of primary biomolecules, as it has been stated, could have emerged as a result of preferential polimerization of monomers of the same sign (Gol'danskii and Kuz'min 1989; Eliel et al. 1994; Plasson et al. 2007; Barron 2009). Even if that is so, *it is still necessary to identify a natural system whereby all characteristics of a probiont could be formed in one place as a result of conjugate elementary stages of a single process.* We discuss these issues in detail in (Tverdislov and Yakovenko 2008; Tverdislov et al. 2012).

One of the considerations, which serve as basis of our long-standing research, is that a bifuraction, whereby *the living nature* on Earth was selected and directed to the path of its co-evolution with the non-living nature, coincided with the formation of two basic conjugated asymmetries, chiral (molecular) and ionic (cellular) (Tverdislov and Yakovenko 2008; Tverdislov et al. 2012). These two asymmetries, which are energetically equal (Tverdislov and Yakovenko 2008; Tverdislov et al. 2012) and informationally complemental for the biological systems, comprised the primary basis and «escort» of biological hierarchies, which defined the formation of unicellular and multicelluar organisms, biodiversity and, hence, an evolutionary trajectory of the biosphere. A characteristic property of the proposed model is that it incorporates mechanisms, by way of which the initially thermodynamically non-equilibrium precursors of living cells originated. In the classic models the initial forms of life, such as Oparin's coacervates, Fox's proteinoids, clay particles or crystalline structures, are supposed to have been in the state of thermodynamical equilibrium, and evolutionary development was launched by an external action that further maintained a «stable non-equilibrium» (Rutten 1971; Shapiro 1987). The previously proposed mechanisms of the asymmetrical distribution of ions and enantiomers of amino acids and sugars together with the common ideas about the emergence of probionts are based on the properties of heterogenic thermodynamical systems in equilibrium, whereas living cells are essentially non-equilibrium.

In the experiments modelling «Oparin's ancient ocean» or «Darwinian puddles» we experimentally confirmed that the «cold surface layer», coming about as a result of heat and mass transfer between sea surface and the atmosphere, is enriched with potassium and calcium ions (Shapiro 1987; Tverdislov et al. 2007; Tverdislov and Yakovenko 2008; Barron 2009). In the presence of racemate blends of certain aminoacids a thin surface layer is enriched with L-enantiomers with the relative enrichment of up to 3-5%, which is more than sufficient for the realization of the evolutionary advantage of one of the enantiomers. «A thin layer» is not formed under the equilibrium conditions, and thus, ion and enantiomer fractionation cannot take place. If a solution, on its surface, contains a rarefied monolayer of abiogenically formed phospholipids, then after the rupture of air bubbles the newly formed lipid vesicules are filled with a solution from the surface layer and enriched with potassium and calcium ions as well as with L-aminoacids (Tverdislov et al. 2007; Tverdislov and Yakovenko 2008). Similar formations can be regarded as cell prototypes, capable to serve as reactors possessing free energy. It may be inferred that from here on starts the evolutionary development of autowave processes and Eigen's cycles, which became subject to selection during the initial biological evolution.

Note that enantiomers of biologically significant chiral compounds can function as a logical element, or as an informational switch, not just at the level of simple «yes/no» coding, but at the level of recoding of a «sense bearing» signal (Fig. 5) – in the presence of a code and a decoding device.

The depicted symbolic representation of informational non-identity of enantiomers may have a direct impact on the problem of «the chiral purity of the biosphere». An established opinion that «the chiral purity of the biosphere» is identical to homochirality at the level of monomers (Kuz'min et al. 1989) is no longer appropriate. With respect to aminoacids the principle of «chiral purity» must be extended to participation of D-isomers in the regulation of the important stages of ontogenesis, while in the past a classical interpretation of this principle included only L-isomers in ribosomal protein synthesis. D-aminoacids (asparagine, serine) are sent to the «managerial» level, regulating the key hormonal and morphogenetic processes (Tverdislov et al. 2011). This is the first example of the functioning of a sign of chiral symmetry in structural and functional organization, inherent, as it appears, to all living organisms on Earth.

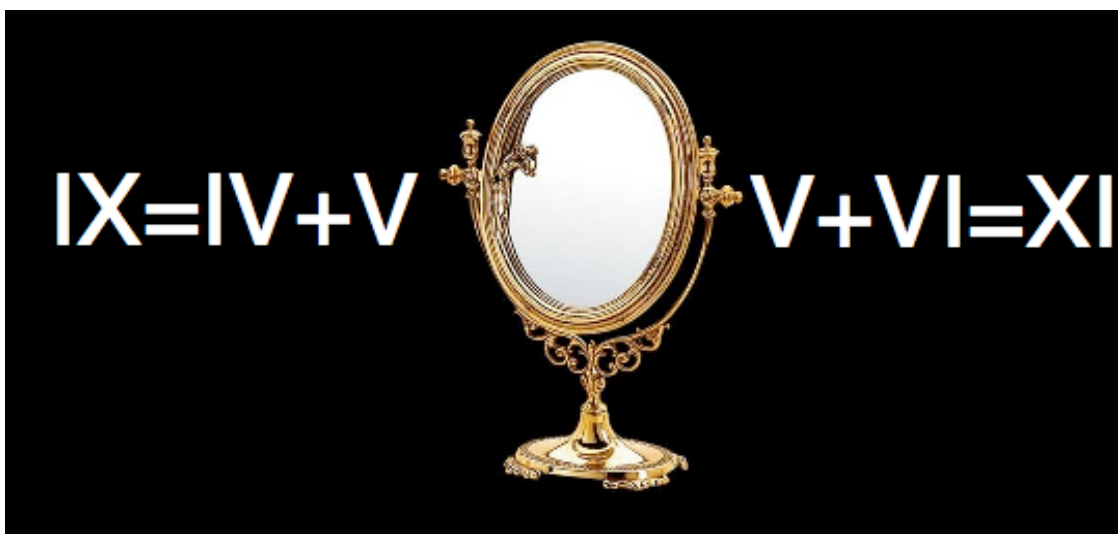

Fig. 5. Example of a chiral pair, in which enantiomorphs possess the same amount of information, but have different sense if code is the same for both parts.

The notion of chirality is closely connected to the notion of complimentarity; there is a stamp and a print. Based on the stereospecificity (no less than three corresponding contact centers without a symmetry axis) the principle of complimentarity is implemented on more than one level of molecular biological hierarchy, including the matrix storage of biological information, replication, transcription and ribosomal synthesis, and further the processes of enzyme catalysis, reception, membrane transport and others. Chiral elements from several structural and functional levels are necessarily engaged in all of the mentioned processes. Only in chiral structures does the principle of complimentarity acquire biological expediency. As it is shown below this property is revealed during intermolecular interactions in the heterochiral systems DNA-protein.

## 3. L/D stratification in molecular biological systems

*Desoxyribose in DNA and ribose in RNA are D-isomers, whereas the constituent nucleotides, which form side groups of a polymeric chain, reside mostly in the left form (gauche conformation). But the DNA double helix is right-handed.* Subsequent supercoiling, inherent to DNA semiflexible polimeric chains, is of topological nature and always reveals itself in «left» twists of the «right» double helices of circular molecules, such as those in bacteria (Waigh 2007; Nelson et al. 2008; Liljas et al. 2009). A coil of a supercoiled double-chain DNA is formed as a result of the rotation of a nucleic acid double helix in the direction opposite to that of

rotation of individual chains of the right-handed double helix. ***Negative supercoiling may lead to the formation, in certain parts of DNA, of a left-handed helix***, i.e. a transformation of B-form to Z-form. Left-handed supercoiling permits the unwinding of the right-handed double helices of DNA, which are necessary for the processes of replication, transcription and recombination.

As a whole, we ***observe a regular alternation of chirality sign – D-L-D-L – during the transition to a higher level of structural and functional organization of DNA.***

A similar alternation of chirality sign is further observed in the sequence of protein synthesis: ribosome-synthesized **popypeptide protein chains are formed from L-aminoacids, whereas the pivotal universal secondary structure, alpha helix, is always right-handed**. As for the energetically favorable structures, formed by an ensemble of secondary and suprasecondary structures, chirality is observed in supercoiled alpha helices in fibrillar proteins. In the Crick-postulated model of supercoiled alpha helix ***two alpha helices are twisted around each other, forming a left superhelix*** with a period of about 140 A (Brändén and Tooze 1999). Supercoiled alpha helical structures were found in fibrillar proteins – alpha-keratin, tropomyosin, paramyosin, and in the light chain of meromyosin (Schulz and Schirmer 1979; Liljas et al. 2009). Short pieces of the suprasecondary structure were found in globular proteins, which contain parallel and anti-parallel alpha helices. Note that the relative direction of alpha helices is not important, because an alpha helix is right-handed in both directions. The known examples of an almost linear packaging of helices include hemerythrin, the coat protein of tobacco mosaic virus, bacteriorhodopsin, the coat protein of bacteriphage fd and tyrosyl tRNA synthetase (Schulz and Schirmer 1979). In these or similar structures a tight packaging of side groups is feasible without distorting alpha helices**.** This stable configuraion of two chains can be unobstructedly prolonged **if two right-handed helices form a left-handed superhelix** with a straight contact axis. Again, in this case we observe a topological correspondence of two chiral structural levels, which change their chirality signs in a way similar to DNA supercoiling, but only with a phase shift. In addition, as Schultz and Shirmer note, supercoiling of alpha helices is energetically favored due to the fact that the side group packaging contributes to formation of additional van der Waals contacts between alpha helices. If interacting side chains are hydrophobic, then a decrease in free energy of that structure will be especially effective, because side chains,

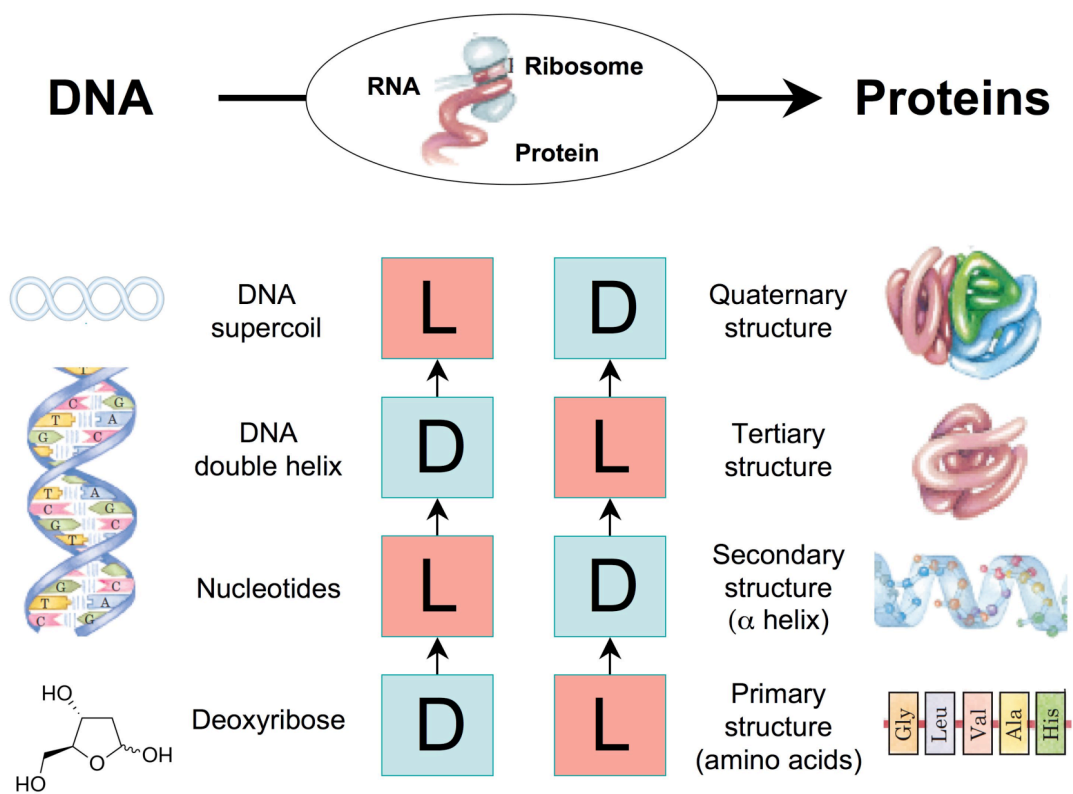

Fig. 6. Hierarchy of chiral structures in DNA and proteins. Left and right columns form chiral pair of a higher level, constituting an achiral invariant.

positioned along the superhelix axis, will be isolated from the contact with water molecules.

Intracellular **supramolecular structures, such as microfilaments in muscles or in cortical layer, as well as cytoskeleton microtubules, are right-handed helices** formed by globular proteins such as actin and alpha- and beta-tubulin (Schulz and Schirmer 1979). **The sequence of chirality sign alternation in the structural and functional hierarchy of protein structures is similar to that we observed in DNA: L-D-L-D. A phase shift is obvious because the protein hierarchy starts from L-aminoacids, and the nucleotide hierarchy starts from D-carbohydrates (deoxy)riboses.**

In principle, the premises of the proposed paradigm prompt an answer to the question about the basis for precisely this relation between chiral hierarchies of nucleic acids and proteins. Probably the answer is in the complimental principles of chirality and complimentarity in the chain DNA-RNA-protein. It appears that **the mechanism of complementary interactions that connects DNA and proteins at the level of RNA and ribosomes provides for another level in the chiral hierarchy of** *informationally determined macromolecular systems (D/L-sequences of DNA and proteins are mirrored). As a matter of fact nucleic acids and proteins were formed in* nature *as linear polimers for the purposes «to build», in the course of early evolutionary adjustment, a «sensible» chiral pair «DNA - protein». Then the widely discussed archaic «RNA world»* (Baserga and Steitz 1993) *can be viewed as a replicative gear, an initial «drive belt» in conjunction with chiral blocks of DNA and proteins. Indeed, it is known that the oligomeric RNA structures are structurally very labile. Then these blocks, enclosed in «chiral forms», could be filled with diverse nucleotide and aminoacid «content». It is natural to suppose that the proposed construction became a primary stable block, connected by chiral couplings, which became a «cornerstone» of modular evolutionary movement of living nature along the hierarchical ladder of structures and functions.* A flow chart of molecular biological L/D positions of chiral compounds discussed herein is presented in Fig. 6.

It is instructive to develop the «chiral-hierarchical» analogies and consider the formation of homochiral linear polimer chains formed by «left» aminoacids and «right» carbohydrates. In linear amylose, a compound of starch, residues of alpha-D-glucopyranose are connected with each other by α-1,4-glycosidic bonds. Oxygen bridges are formed between the glycosidic hydroxyl of the first atom of an alpha-D-glucopyranose molecule and the alcohol hydroxyl of the fourth atom of another molecule of alpha-D-glucopyranose. Residues of alpha-D-glucopyranose reside in the Boat conformation, which contributes to the coiling of polyglycoside chain.

Indeed, one of the first discovered (in 1943) helical structures formed by biopolymers was the *left-handed helix of amylose*, that winds around iodine molecules in the known complex of iodine and starch (Imberty et al. 1988) (Fig. 7). The number of residues per pitch is equal to 6, size of a pitch is 0.8 nm, its diameter is about 14 nm. Thus the rule of chirality sign change is again confirmed: **the «right» D-glucose forms the «left» helix of amylose.** Molecules of amylose, as well as of other linear polysaccharides can interact with each other forming non-helical secondary structures with reciprocally twisted polysaccharide chains, where chains are extended in the same or in opposite directions.

The presented regularity of chiral conjugation of strata in the hierarchical systems may be compared to quantum-mechanical systems, where a structure is governed, for example, by the Pauli principle (Penrose 2005). However, in macroscopic systems one has to consider, so it seems, tendencies and trends rather than rigorous restricting rules. That is why a table of chiral structures may have exceptions. Thus we have evidence that the DNA superhelix of archaea is right-handed contrary to the left-handed superhelix in bacteria.

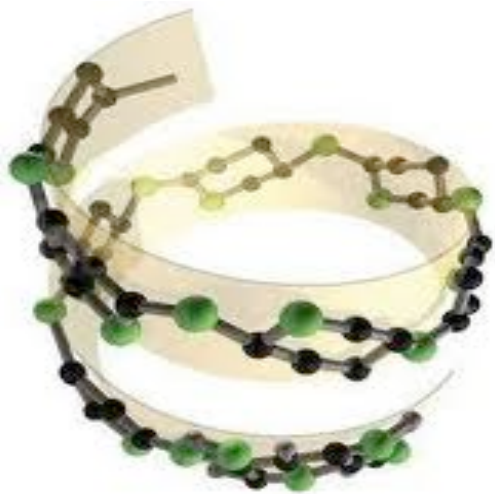

Fig. 7. «Left» spiral of amylose, formed from molecules of D-glucose (Imberty et al. 1988).

There exists another, a very important, but poorly investigated and little-discussed problem of the biological significance of homochirality of phospholipids, which constitute cellular membranes. It is related to the problems of structure formation in proteolipid membrane complexes and to the problems of interaction of chiral elements of proteolipid systems responsible for energy transforming, as well as of enzyme, transport and receptor membrane systems. Indeed, the second carbon atom of glycerol in phosphoglyceride is asymmetrical and has L-configuration, because glycerol originated from L-glyceraldehyde (Nelson et al. 2008). On the whole, homochirality of phospholipids is connected with the enzymatic nature of phospholipid synthesis and degradation, and, in particular, with the functioning of enantiospecific phospholipases. One may presume that phospholipid homochirality serves as an additional order parameter in the formation of a lipid bilayer structure and in the control of its phase state. It is to be remembered that integral membrane proteins have a relatively high proportion of transmembrane alpha-helices, which form mostly left-handed superhelices; these, in turn, can lead to a variation in the affinity of interactions of D and L-helical protein structures with L-homochiral phospholipids.

Throughout decades, an important task of the physics of biopolimers was the problem of gradation of structural levels of macromolecules. The levels of primary, secondary, tertiary and quaternary structures have been distinguished. The principle of chiral stratification details and specifies the accepted classification, namely the physical discreteness of the levels and the peculiarities of their interactions. In subsequent physical models of structural transition in macromolecular (including membrane) systems, historically referred to as «jelly-colloid» transitions, transitions «order-disorder» or «spiral-clew» one has to consider this additional «order parameter».

## 4. Active media and molecular machines as chiral structures

The ideas about the chiral order of the complex systems are immediately related to the ideas concerning active media and molecular machines, which represent some of the major achievements in biophysics over the last fifty years (Blumenfeld and Tikhonov 1994; Kondepudi and Prigogine 1998). In themselves, the molecular structures cannot be considered alive until their ensemble becomes capable of transforming matter, energy and information. This transformation is cyclic in its nature. The cyclicity implies a transition from non-biological to biological level, from structural to functional motives. In this manuscript we discuss two types of cyclic systems: molecular machines and active media under the regime of autowave self-organization.

Whereas the laws of thermodynamics were based on the gas laws, then the existence of active media and molecular machines is bound to the condensed phases, namely, to the liquid phase in the case of active media («reaction – diffusion – autocatalysis - non-linear transformer of energy/ matter/ information»), and to the solid or liquid crystal phase in the case of molecular machines. In both cases we are referring to the energy transforming systems. Active media and molecular machines brought about the notion of non-thermal (mechanical/ quantum mechanical) selected degrees of freedom, which cannot originate in the gas phase. Structurally determined as a result of an evolutionary selection, elementary transformers of energy and information became molecular or, in the aggregate, macroscopic biological machines, whose constructions were deposited in the genetic machinery of individuals. This stage signified the transition from non-living to living nature.

Dissipative autowave structures of active media may form various types of symmetry together with their chiral subsystems, and create corresponding selected degrees of freedom. However these structures are short-lived due to the high speed of dissipative processes in a phase devoid of long-living «beams» and «joints» (Blumenfeld and Tikhonov 1994). The situation is different in solid or liquid crystal macromolecular constructions.

The main properties of active media as symmetry generators can be described as follows (Tverdislov and Sidorova 2012).

(i) Active media are characterized by the presence of distributed resources (energy, matter, information); they belong to a unique class of non-linear systems, which can be described in terms of «reaction-diffusion-autocatalysis»; they exhibit a property of self-organization, which reveals itself in the formation of autowave dissipative structures. In the simplest case these are concentric autowaves, propagating from a pacemaker.

(ii) During formation of autowave patterns the number of initial degrees of freedom effectively describing a system decreases (initially the system is a homogeneous autocatalytic space or a space of finite-dimensional autocatalytic oscillators, generators of autooscillations). Selected degrees of freedom define the type of symmetry of the created structures and the trends of their evolution. Essentially, for evolutionary systems this is a transitional path from active media to machines.

(iii) The refractory zone in active media excludes interference, and in a system of concentric autowaves a pacemaker with the highest frequency occupies the entire reaction space, «eating up» the slower neighbors without a change (importantly) of the system's symmetry type. This kind of competition illustrates a kinetic criterion of selection: a space is occupied by the system that transforms free energy more efficiently.

(iv) In homogeneous active media propagation velocity, length and shape of an autowave are constant and do not depend on the initial and boundary conditions. Trivial concentric structures in non-homogeneous active media may be transformed into chiral patterns. When a plane autowave encounters a non-homogeneous slope-like barrier, then, depending on the slope inclination, it is transformed into a spiral with a chirality sign, which is a right or left reverberator. A model of symmetry change in non-homogeneous active media was proposed by M. Poptsova (personal communication, 2004).

(v) Interaction of chiral processes with chiral structures in a lateral (quasi two-dimensional) system may also change the correlation of mirror components in the developing processes. It is essential that this kind of process is irreversible.

(vi) Due to peculiarities of their formation, spiral autowaves (reverberators) have minimal wavelength and maximal competitiveness («kinetic perfection») in the struggle for reaction space with autowave patterns of other symmetries.

(vii) In the course of autowave self-organization active media may conjugate processes of commensurate spatial and temporal scale. Similar systems automatically become coherent.

The irreversibility of chiral stratification of macroscopic and molecular structures is based on different mechanisms. A change of the chirality sign of an autowave reverberator is connected with the necessity of its complete destruction and consequent reassembly, which is impossible in a spontaneous way; in order to form its antipode a structure with a given type of symmetry must disappear. A macroscopic homochiral structure at the level of active media cannot spontaneously change its chirality sign (for example, L for D), and its symmetry rank cannot decrease (e.g., transformation from a spiral to a circle is not feasible). Spontaneous reduction is impossible. The qualitative interpretation is that the lower is the symmetry the more complex is the structure. None of the structures can exist in a completely symmetrical system (an isotropic infinite medium).

The chiral irreversibility is lost at microlevels. At the molecular level chiral compounds undergo an isoenthalpic spontaneous process of racemization, where entropy increases. ***In molecular systems resistance of enantiomers to racemization is provided for their incorporation in macromolecular hierarchies.***

Thus, in the double-chained DNA, a change in the primary structure of D-glucose by its L-isomer will lead to the loss of complimentarity, breakage of several base pairs and separation of the chains due to the steric incompatibility. This effect can be easily demonstrated with a molecular constructor or computer simulation. An impact of the upper layer of the structural hierarchy on the double chain can be demonstrated by the effect of tension of the left-handed superhelix of the circular DNA on the winding-unwinding processes (Waigh 2007). Similarly, one can consider the role of protein «body» in the processes of enzyme catalysis, mechanochemical transformations and membrane transport.

Thermodynamics as a branch of science does not deal with symmetries directly, but encounters those in self-organizing dissipative systems – active media and biological machines, both macroscopic and microscopic. It should be mentioned that Prigogine's rule "regulates" transformations in systems, which possess symmetrical elements: in the course of the unfolding processes a linear thermodynamic system cannot lower its symmetry rank, i.e. it cannot transform its structure in the direction from maximally symmetric infinite isotropic media towards maximally anisotropic, or vector medium. This rule is not applicable to non-linear systems such as active media or machines. At the level of thermodynamics some systems, which are close to the state of thermodynamic equilibrium, can be considered as linear and must be locally reversible. However kinetically these become irreversible due to the presence of chiral objects. A chiral element as a construction block of a machine makes a system – which is close to the state of thermodynamic equilibrium – a non-linear system. Non-linearity, inherent to the (biological) machines, is defined, in particular, by the presence of chiral elements.

*Active media generate symmetries; machines use them.* In the general case *a machine is a device (construction) that is capable of transforming energy in a cyclic manner, performing "effective" work due to the presence of "selected mechanical degrees of freedom" (translational, rotational), which kinetically divide work and dissipation*. Naturally, isotropic or chaotic structures are not capable of shaping machines up. *For realization of cyclicity as its principal temporal characteristic a machine must be symmetrical in construction, but for the movement along the loop of the cycle in the «right» direction it requires an asymmetrical element with the property of a «faucet» or a «latch»* (Blumenfeld and Tikhonov 1994). Thus, a well-known element of clockwork, ratchet and pawl, which was presented in detail by Feynman (Feynman et al. 2006), is also a chiral object. In principle *any machine is a chiral object, an enantiomorph*. For that reason a reversion of the cycle requires the change of a chirality sign of the faucet element (L/D change by convention), an interplay between the «entrance» and the «exit». An electromotor becomes a current generator, a refrigerator becomes a heat pump, a membrane proton pump and Na-pump turn into ATP-synthase (Romanovsky and Tikhonov 2010). Another *principally important peculiarity of molecular machines is the cooperative* *participation of secondary, tertiary (and quaternary, if supramolecular complexes are present) macromolecular structures in cyclic work, which, to a high degree, is ensured by the hierarchical conjugation of corresponding «chiral» degrees of freedom.*

Note that autowave self-organization of biological machines is also possible at the supracellular (supraorganismal) level, if system elements are connected through "diffusive degrees of freedom".

## 5. Self-organization in the hierarchy of active media as one of the fundamental driving forces of the evolution of biosphere

Symmetries and chiral asymmetries create a regular medium that generates selected degrees of freedom and types of movement on all structural and functional levels of the biosphere. We present a point of view whereby chirality, as a dialectic manifestation of dualism, rather than becoming an artefact in the evolutionary process, developed into one of the important instruments of biological evolution. The «chiral purity of the biosphere», in our opinion, consists not in the exclusion of enantiomers or enantiomorphs of one sign from the evolutionary process, but in their specialization, e.g. separation according to function and hierarchical level. A dualistic model of evolutionary development herein proposed offers a novel approach to the processes of bifurcational, and, consequently, saltatory development of populations, organismal communities and the biosphere as a whole.

What follows is a discussion of the meaning of chiral stratification. The resulting progressiveness of biological evolution seems to be obvious, however, scientists have difficulty defining the driving forces of evolution, its quantitative and even terminological criteria. It has become common to use the term "complexity", which does not easily yield to denotation. Progressive evolution is interpreted as an increase in complexity (Adami et al. 2000). It appears that one of the more common and illustrative approaches to the definition of complexity should comprise the notions of symmetry and asymmetry, and, in particular, that of chirality. An adequate biophysical model of the evolution of the biosphere may, in particluar, rely on the ideas about self-organization of chiral, hierarchically conjugated structures. According to this approach «an increase in complexity» is equivalent to a decrease in the degree of symmetry, a decrease in the number of selected degrees of freedom. The transition from stochastic formations to molecular machine constructions occurred in the course of prebiological evolution and at the initial stages of biological evolution; such transition is deemed impossible in linear thermodynamic systems where an increase in the level of symmetry towards an isotropic medium is only permitted (Kelvin 1904). During this process a multiplication of chiral hierarchies implies an increase in the general «complexity» of biological systems at the expense of the choice of the types of symmetry during chiral selection.

As in the living nature, the major biophysical mysteries of (pre)biological systems lie at the boundary of microscopic and macroscopic worlds. In a similar way, the chiral biological structures are natural to divide into two classes: the molecular structures, whose chirality is connected with the peculiarities of carbon compounds and is determined in the process of biosynthesis, and macroscopic dissipative structures. In achiral molecular systems macroscopic chiral structures can emerge at the supramolecular crystal or liquid crystal level due to their asymmetrical packaging, or as a result of autowave self-organization in active media.

The above-presented sequence of chiral transformations at the molecular biological level is not an end in itself, but one of the important manifestations of the evolutionary development of the biosphere. As we already discussed, ***chirality of carbon compounds*** is the prime cause of formation of a discrete hierarchical structure of the evolving biosphere, whose stratification is fixed by a non-reversible change of the chirailty sign of an enantiomorph during transition to the next level. The mechanical degrees of freedom in macromolecules, associated with chiral structures, are deterministically interconnected, so that macromolecular structures become cooperative systems. This is, we believe, the primary finding of the living nature as it is connected with a «modular» (as opposed to step-wise, additive) character of the biological evolution, based on the initial chirality of carbon compounds.

As to the meaning of «the chiral degrees of freedom» in molecular structures (this aspect is usually neglected in physics literature), these define the unique properties of chiral systems, which are related to the collective character of transitions between adjacent structural levels: a change of chirality sign on one hierarchical level necessarily leads to the to change of chirality sign on adjacent levels. We are convinced that thus the living nature prevents the destructive action of heat noise at the molecular level.

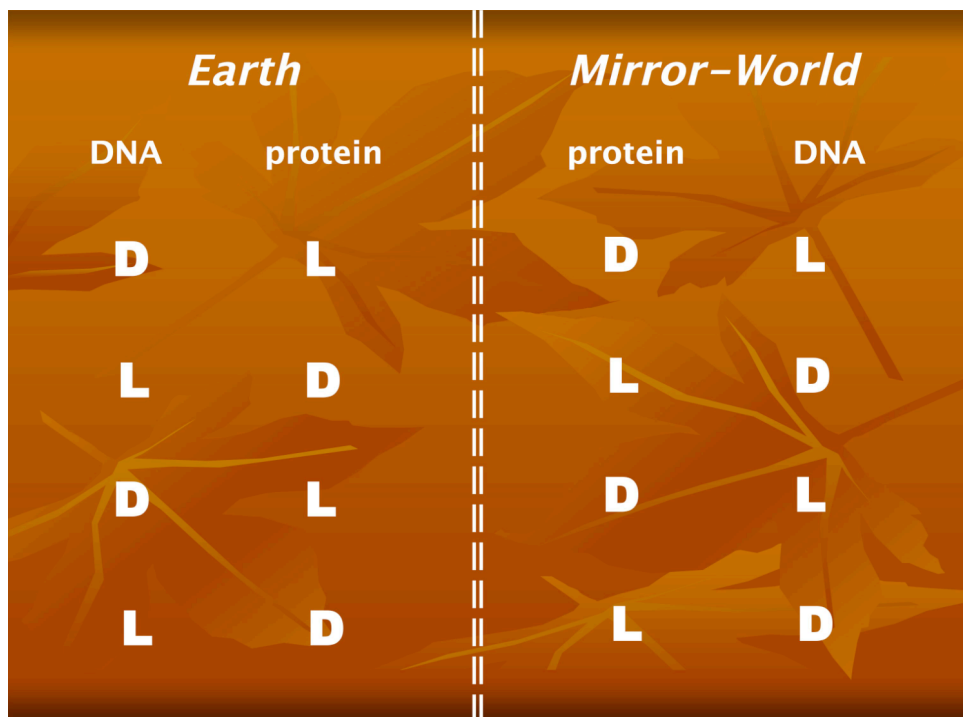

Fig. 8. In the mirror-world the module, comprising hierarchies of chiral structures of DNA and proteins, will look the same, since every element in the diagram will be twice mirror transformed: as an individual chiral structure and as a planar chiral pair.

DNA and a protein, as heterochiral hierarchical structures, form a chiral pair, which constitute an achiral invariant, that is, as we suggest, the highest and universal unit of molecular biology, the basic module of the evolutionary development of the biosphere. (Fig. 8). The conjugated block "DNA-protein" constitutes a chiral shell that, in the course of an evolutionary development of a taxon, can be filled by various combinations of four nucleotides and twenty aminoacids. Note that this block possesses a symmetry axis, and hence it is achyral. Also, the considered module is an invariant; behind the looking glass it will look the same, since every element in the diagram will be twice mirror transformed. The achiral invariant is capable to enter the macroevolutionary process as a whole entity. At this level the direct influence of asymmetric carbon is depleted and the collective struggle of deterministic structures against the destructive action of heat noise is ended. With this invariant (presumably) molecular biology ends and cell biology begins. A transition occurs from molecular complementarity and self-assembly to autowave self-organization of active media.

It should be noted that the above-described invariant is also self-sufficient and cyclically closed. The DNA together with the RNA-ribosome system form a cyclic device, a machine. Through a network of complementary interactions this machine, as if it were led by a drive gear, conjugates the processes of translation, transcription and ribosomal syntheis, i.e. both L/D columns of DNA and proteins (Fig. 8). Specialized proteins, in turn, regulate processes with the participation of DNA.

At the next stage of building up a presently qualitative model of chiral hierarchies of biological self-organizing structures the key question is the system transition in hierachy: how do chiral macromolecular and chiral autowave constructions conjugate in biological systems? It seems that the levels of macromolecular chiral self-organization and autowave self-organization are conjugated in the cells and tissues through participation of regulatory systems of the third type, for example, the ionic systems. It is important that macroscopic autowave intracellular and organismal processes are channeled through the heterogenic biological structures (for instance, through the membrane), which govern ionic flows. Ionic pumps and ionic channels, membrane systems of simport and antiport as well as receptors are all chirally determined protein molecular machines.

A well-known example of the conjugation of molecular and autowave processes is the nerve impulse propagation, which represents a system with feedback: an autowave of ion redistribution between a neuron and a medium by way of electrostatic potentials causes intracellular reconstructions in protein ionic membrane channels, and the changes in the conduction of the potential-sensitive channels lead, in turn, to the electric potential front movement. Another example is the launch of mitotic processes in a fertilized egg-cell by the ionic «calcium» autowave that activates membrane ion transporting systems (Gilbert 2003). Examples also include the processes of electro-mechanical autowave propagation in the heart muscle during contractions.

Evidently, the genesis of conjugation of these two types of hierarchies is closely connected with the problem of emergence of the living cell precursors. We return to the idea of conjoined origination of two fundamental asymmetries, chiral and ionic, in non-equilibrium surface structures of the ancient ocean (Tverdislov and Yakovenko 2008). Chiral asymmetry is localized in molecular structures, ionic asymmetry is distributed in intracellular and supracellular spaces. Furthermore, these two fundamental asymmetries are related by their common origins, they are spatially interlaced. Spatially and functionally, the ionic asymmetry is a shell for chiral asymmtery because sodium-potassium, magnesium-calcium and proton asymmetries support a cellular reactor in a stable non-equilibrium state where, along the flow of energy, sign-alternating chiral DNA-protein blocks are reproduced.

### 6. Hierarchies and fractals

Presently an opinion prevails that non-equilibrium systems in non-linear regions can develop non-deterministically via a succession of bifurcations. Non-linearity of systems, which are far removed from the state of thermodynamic equilibrium, shows not only in the peculiarities of the development of such systems with time, but also in the peculiarities of their spatial structure evolution (Dyson 1988). Classical interpretation of the known bifurcational curves, describing the evolution of a system, usually employs «order-disorder» as the principal terms. We think it reasonable to include into the list of characteristics of «evolution through bifurcations» the change of a class and a sign of symmetry for chiral objects. It is very plausible that *there exists a certain general principle of «conservation» in «a succession of symmetries» during transition of a chiral system through hierarchical levels.* Usually symmetries of the same type may coexist at the same or adjacent levels, but only symmetry of some type is disconnected by the sign of chirality. When it concerns spiral formations, the «left» elements ascending to the next level form the «right» macrostructures, and the «right» macrostructures form the «left» megastructures, etc. At one level a relatively simple hierarchy of spirals is replaced with a sequence of symmetries of a higher rank. *Types of symmetry multiply in lateral interactions and have a tendency to change their sign during the transition to the next level.*

Consideration of natural and model hierarchical systems with chiral subsystems has unexpectedly raised an issue about insufficiency and even non-adequacy of the description of such systems with the known fractal models. This is due to the conservation of symmetry sign in the chiral fragments, which is a key property of multiscale self-similarity (Mandelbrot 1983). Non-adequacy of fractal description of real hierarchical systems lies in the kinetic reversibility of the transition between levels: the «complexity» of fractal invariants does not change irreversibly in multiscale self-similar structures. A «latch» is absent. This does not necessarily mean that there is no interlevel homochirality in nature. It is likely that such self-similarity is possible, but it never becomes a condition for determinism and stability of strata.

## Conclusions

We have for the first time noted the presence of L/D-sign-alternating structural hierachies in DNA and proteins, which form a chiral pair that in turn comprises an achiral invariant; the latter defines the modular-and-saltatory character of biological evolution. As a whole, the presented regularity demonstrates unity and succession of evolutionary processes in non-living and living nature based on dynamic development of chiral molecular and macroscopic autowave systems.

To summarize, *any chiral system, possessing free energy, tends to spontaneous formation of a new, higher structural level with the same type of symmetry but with an opposite sign of chirality and on an enlarged scale. The resulting hierarchy of conjugated sign-alternating chiral structures:*

*- kinetically stabilizes them by inhibiting spontaneous racemization;*

*- forms a conjugated system with selected degrees of freedom, which makes the work of biological machines possible;*

*- in living systems macromolecular and autowave chiral structures are conjugated as a consequence of their channeling through the structures of chiral biological machines;*

*- defines the vector of the general development of a system in the direction of an upper, «open-ended» hierarchical level.*

## Acknowledgments

The author thanks M.S. Poptsova for invaluable help in preparation of this manuscript. The author is also grateful to L.V. Yakovenko for longstanding co-operation, and to S.V. Stovbun, T.A. Preobrazgenskaya, A.E. Sidorova, A.N. Tichonov, A.A. Ivlieva for fruitful discussions.